\documentclass{PoS}

\usepackage{comment}
\usepackage[utf8]{luainputenc}
\usepackage{array}
\usepackage{booktabs}
\usepackage{blindtext}
\usepackage{support-caption}
\usepackage{subcaption}

\def\lsima{$\; \buildrel < \over \sim \;$}
\def\lsim{\lower.5ex\hbox{\lsima}}

\title{The Rate Of Short Duration Gamma-Ray Bursts In The Local Universe}

\ShortTitle{SGRB Rate in the Local Universe}

\author{\speaker{Soheb Mandhai}$^{,a}$ Nial Tanvir$^{a}$ Gavin Lamb$^{a}$ Andrew Levan$^{b,c}$
and David Tsang$^{d}$,\\
\llap{$^a$}Department of Physics and Astronomy, University of Leicester, University Road, LE1 7RH, U.K.\\
\llap{$^b$}
Department of Astrophysics/IMAPP, Radboud University, Nijmegen, The Netherlands\\
\llap{$^c$}
Department of Physics, University of Warwick, Coventry, CV4 7AL, U.K.\\
\llap{$^d$}Department of Physics, University of Bath, Claverton Down, Bath, BA2 7AY, U.K.\\
E-mail: \email{sfm13@leicester.ac.uk}, \email{nrt3@leiceseter.ac.uk},
\email{gpl6@leicester.ac.uk}, \email{A.Levan@astro.ru.nl}, \email{D.Tsang@bath.ac.uk}}

\abstract{The binary neutron star merger responsible for the gravitational wave event, GW170817, strengthened the merger association with short-duration gamma-ray bursts (SGRBs) following the detection of the SGRB counterpart, GRB 170817A. Here we consider the constraints on a population of local gamma-ray bursts with moderately short duration ($T_{90}<4$ s) and within $d < 200 $ Mpc, that may have originated from similar compact binary mergers. Using well localised gamma-ray bursts from $\sim14.5$ years of \textit{Swift}/Burst Alert Telescope monitoring, we find 
no events with high likelihood of being in this distance range, and place
an upper limit for the all-sky rate of such events of $<4\ y^{-1}$.
For \textit{Fermi}/Gamma-ray Burst Monitor (GBM) and \textit{CGRO}/Burst And Transient Source Experiment (BATSE) detected bursts, where the localisation has considerably larger uncertainties, we cross-correlated with 2MASS Redshift Survey galaxies at $d<100$ Mpc, obtaining a weaker constraint of $< 12\ y^{-1}$. A separate correlation search from the GBM and BATSE bursts for giant flares originating from soft gamma-ray repeaters in nearby galaxies ($d < 11 $ Mpc) yields an upper limit of $<3\ y^{-1}$.  }

\FullConference{The New Era of Multi-Messenger Astrophysics - Asterics2019\\
		25 - 29 March, 2019\\
		Groningen, The Netherlands}

\begin{document}
\section{Overview}
In light of GRB 170817A that accompanied a gravitational wave detection of a merging neutron star binary at d$\sim40$ Mpc, we consider the observational constraints on a nearby population of low-luminosity short-duration gamma-ray bursts (SGRBs), where we extend the definition to include bursts with $T_{90}<4$ s.

Compact binaries consisting of a neutron star paired with either a black-hole or another neutron star may receive natal kicks from the supernovae during formation \cite{bray2016}. Sufficiently high kick velocities 
can eject a binary  from its host galaxy. Thus, although the majority of SGRBs (including GRB\,170817A) occur within the bodies of their host galaxies in projection \cite{Fong2013}, binaries of this nature may merge at large separation from their host and hence their distance from us will not be determined.

We consider two 
approaches 
to identifying potential nearby host galaxies associated with catalogued SGRBs, allowing for the possibility of such kicks.
In Section \ref{s:sgrb-loc}, we determine the number of detected Neil Gehrels Swift Observatory (\textit{Swift}) bursts that have tentative nearby host galaxies that fulfil our criteria. In Section \ref{s:sgrb-cor}, we describe a cross-correlation of catalogued SGRBs with galaxies within $100$ Mpc.

\section{Using \textit{Swift} to Localise Short-Duration Gamma-Ray Bursts}\label{s:sgrb-loc}
\textit{Swift} is a dedicated detection and follow-up satellite designed to observe gamma-ray bursts using the on-board Burst Alert Telescope (BAT) and their afterglows with the X-Ray Telescope (XRT) and UV/Optical Telescope (UVOT) \cite{gehrels2004swift}. \textit{Swift} provides 

\begin{figure}[b]
\centering
    \hbox{
    \hspace{0.13\textwidth}
    \begin{subfigure}{0.35\textwidth}
		\includegraphics[scale = 0.45]{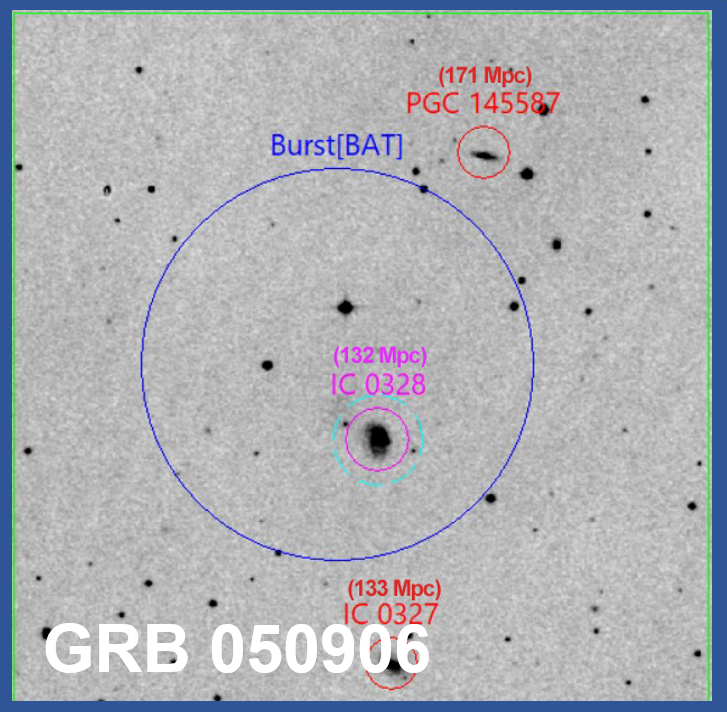}
	\end{subfigure}
	\hspace{0.005\textwidth}
    \begin{subfigure}{0.35\textwidth}
		\includegraphics[scale=0.444,trim={0.04cm 0.03cm 0.045cm 0.05cm},clip]{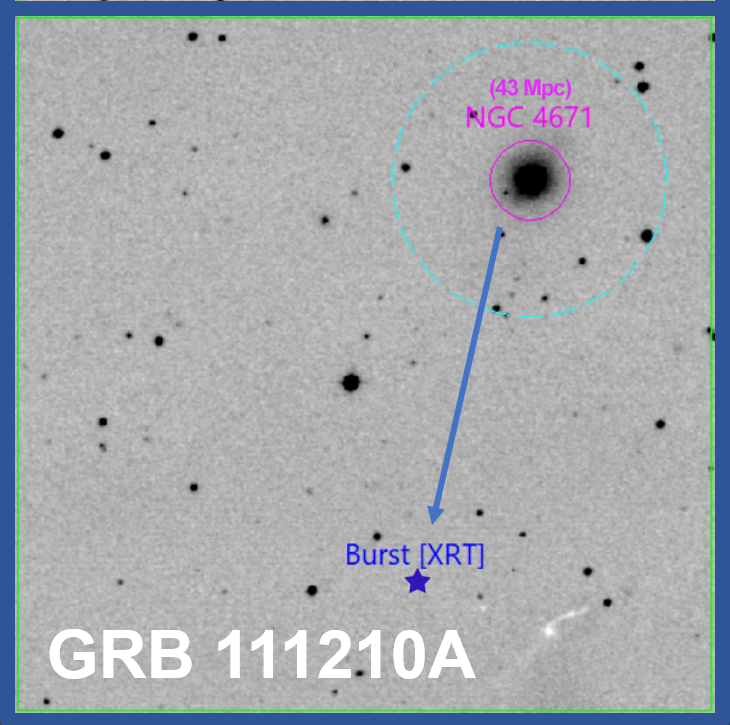}
	\end{subfigure}
		}
    \caption{Potential galactic hosts to SGRB events that are within a projected distance of $200$ kpc from the burst's location. [Left panel] An example case using GRB 050906, where 
    a potential host, IC 0328, appears within the BAT uncertainty region. For this example, no X-ray or optical counterpart was identified.
    [Right panel] An example case of GRB 111210A, where a potential host galaxy (NGC 4671) is found at a displaced distance from the XRT localisation of the burst. In each case, the dashed cyan circle corresponds to a $25$ kpc offset from the candidate host \cite{mandhai2018rate}. }
	\label{fig:sgrb-localisation}
\end{figure}

We searched for {\em Swift} SGRBs that lie within an 
on-sky projection distance of $200$ kpc (this distance corresponds to a long lived binary that has reasonable natal kick of $\sim 100$\,kms$^{-1}$) to galaxies within $d=200$ Mpc from the 2MASS Redshift Survey (2MRS) \cite{Huchra2012}.
Table \ref{tab:sgrb-loced} lists the $10$ bursts out of a total of $\sim150$ {\em Swift} detected gamma-ray bursts with T$_{90} < 4$ s, that fulfill this criteria.
None of these cases can be considered high confidence associations, and the total number  is consistent with the expected rate of chance alignment of background bursts with foreground galaxies.
We conclude an upper limit on the all-sky rate of local SGRBs to be $< 4\ \textrm{y}^{-1}$  . Typical examples of bursts with either BAT or XRT error regions are shown in Figure \ref{fig:sgrb-localisation}.

\begin{table}[]
\caption{
Compilation of {\em Swift} detected bursts that have candidate host galaxies within 200 Mpc from the 2MASS Redshift Survey (2MRS).  For other potential SGRB events that have tentative galactic hosts that are not present in 2MRS but have been discussed in literature are listed below the table break \cite{mandhai2018rate}.}
\label{tab:sgrb-loced}
\scalebox{0.87}[0.87]{
\begin{tabular}{>{\centering}m{1.2cm}>{\centering}m{0.7cm}>{\centering}m{1.5cm}>{\centering}m{2.6cm}>{\centering}m{1cm}>{\centering}m{1.7cm}>{\centering}m{1cm}>{\centering}m{1.3cm}>{\centering}m{1.6cm}>{\centering}m{1.0cm}}
\toprule 
\textbf{GRB}  & \boldmath{$T_{90}$} \textbf{{(}s{)}}  & \textbf{ Angular Separation {(}arcmin{)}}  & \textbf{Closest Galaxy}  & \textbf{Galaxy Type}  & \textbf{Optical Bands (B/R ) {(}mag {)} } & \textbf{J-Band {(}mag{)}}  & \boldmath{$d$} \textbf{{(}Mpc{)}}  & \textbf{ Impact Parameter {(}kpc{)}}  & \boldmath{$E_{{\rm iso}}$} \textbf{}%
\mbox{%
\textbf{{(}$10^{46}$ ergs{)}}%
}\tabularnewline
\midrule 
050906  & 0.26  & 2.0  & IC 0328  & Sc  & 14.0 {(}B{)}  & 12.2  & 132 \cite{HyperLEDAIII}  & 77 $\pm$ 109  & 1.9\tabularnewline
100213A  & 2.40  & 5.4  & PGC 3087784  & S0-a  & 14.7 {(}B{)}  & 11.3  & 78 \cite{HyperLEDAIII}  & 123  & 39.9\tabularnewline
111210A  & 2.52  & 6.0  & NGC 4671  & E  & 13.4 {(}B{)}  & 10.1  & 43 \cite{HyperLEDAIII}  & 76  & 7.5\tabularnewline
120403A  & 1.25  & 4.9  & PGC 010703  & Sc  & 14.4 {(}B{)}  & 12.1  & 133 \cite{HyperLEDAIII}  & 192 $\pm$ 90  & 38.2\tabularnewline
130515A  & 0.29  & 8.5  & PGC 420380  & S0-a  & 16.0 {(}B{)}  & 12.3  & 73 \cite{Cosmicflows3}  & 180  & 28.4\tabularnewline
160801A  & 2.85  & 6.7  & PGC 050620  & Sa  & 15.2 {(}B{)}  & 12.4  & 59\cite{HyperLEDAIII}  & 115  & 10.7\tabularnewline
181126A & 2.09 & 39.8 & NGC 3125 & E  & 13.00 {(}B{)}  & 11.3 & 14.9\cite{EDD09}  & 173 & 1.9\tabularnewline
\midrule 
070809  & 1.30  & 2.0  & PGC 3082279 \cite{070809GCN}  & Sa  & 16.3 {(}B{)}  & 13.5  & 180 \cite{helou1991nasa}  & 105  & 64.4\tabularnewline
090417A  & 0.07  & 4.4  & PGC 1022875 \cite{090417GCN}  & S0-a  & 15.9 {(}B{)}  & 13.4  & 360 \cite{helou1991nasa}  & 461 $\pm$ 292  & 24.5\tabularnewline
111020A  & 0.40  & 2.3  & FAIRALL 1160  & Sa  & $\sim14$ {(}R{)}  & 11.7  & 81 \cite{tunnicliffe2013nature}  & 54  & 9.4\tabularnewline
\bottomrule
\end{tabular}}
\end{table}

\section{Correlating \textit{Fermi}/GBM and CGRO/BATSE Gamma-Ray Burst locations with 2MRS Galaxies}\label{s:sgrb-cor}
In comparison to \textit{Swift}, the localisations from \textit{CGRO}/BATSE and \textit{Fermi}/GBM are considerably poorer but their fields of view are 
greater. As such, the joint BATSE+GBM population of observed gamma-ray bursts with T$_{90} < 4$ s is larger at $\approx800$.
We performed a spatial
cross-correlation between these bursts and 2MRS galaxies 
(as described in \cite{mandhai2018rate}, see also \cite{tanvir2005origin}) for the purposes of constraining the fraction of bursts that arise in galaxies within $100$ Mpc. This yields a maximum upper limit on the all-sky rate of $\lsim 12 \ \textrm{y}^{-1}$.  An additional correlation using GBM and BATSE bursts with a T$_{90} < 1$ s 
to constrain the soft gamma-ray repeaters (SGRs) giant flare rate in nearby galaxies ($d < 11 $ Mpc),
yielded a rate of $<3\ y^{-1}$.

\section{In the Era of LV-O3}
So far, during the third LIGO/Virgo science run (LV-O3) there has been one detection (S190425z) with a high probability of originating from a neutron star binary merger within $200$ Mpc. There were no detections of an accompanying gamma-ray transient reported, despite the localised field being observable by \textit{Fermi}    \cite{Fletcher2019gcn}, and \textit{Integral} \cite{Carillo2019gcn}. 

\section{Acknowledgements}
 We acknowledge 
 use of the 
 HyperLEDA; Extragalactic Distance Database; the NASA/IPAC Extragalactic Database.
 S.F.M. is supported by a PhD studentship funded by the College of Science and Engineering at the University of Leicester; G.P.L. is supported by STFC grants; N.R.T. and A.J.L. acknowledge support from ERC Grant 725246 TEDE.

\bibliographystyle{jhep}
\bibliography{biblio}

\providecommand{\href}[2]{#2}\begingroup\raggedright\begin{thebibliography}{10}

\bibitem{bray2016}
J.~C. {Bray} and J.~J. {Eldridge}, \emph{{Neutron star kicks and their
  relationship to supernovae ejecta mass}},
  \href{https://doi.org/10.1093/mnras/stw1275}{\emph{\mnras} {\bfseries 461}
  (2016) 3747}.

\bibitem{Fong2013}
W.~{Fong}, E.~{Berger}, R.~{Chornock}, R.~{Margutti}, A.~J. {Levan}, N.~R.
  {Tanvir} et~al., \emph{{Demographics of the Galaxies Hosting Short-duration
  Gamma-Ray Bursts}},
  \href{https://doi.org/10.1088/0004-637X/769/1/56}{\emph{\apj} {\bfseries 769}
  (2013) 56}.

\bibitem{gehrels2004swift}
N.~Gehrels, G.~Chincarini, P.~Giommi, K.~Mason, J.~Nousek, A.~Wells et~al.,
  \emph{The swift gamma-ray burst mission}, {\emph{The Astrophysical Journal}
  {\bfseries 611} (2004) 1005}.

\bibitem{Huchra2012}
J.~P. {Huchra}, L.~M. {Macri}, K.~L. {Masters}, T.~H. {Jarrett}, P.~{Berlind},
  M.~{Calkins} et~al., \emph{{The 2MASS Redshift Survey -- Description and Data
  Release}}, \href{https://doi.org/10.1088/0067-0049/199/2/26}{\emph{\apjs}
  {\bfseries 199} (2012) 26}.

\bibitem{mandhai2018rate}
S.~Mandhai, N.~Tanvir, G.~Lamb, A.~Levan and D.~Tsang, \emph{The rate of
  short-duration gamma-ray bursts in the local universe}, {\emph{Galaxies}
  {\bfseries 6} (2018) 130}.

\bibitem{HyperLEDAIII}
D.~{Makarov}, P.~{Prugniel}, N.~{Terekhova}, H.~{Courtois} and I.~{Vauglin},
  \emph{{HyperLEDA. III. The catalogue of extragalactic distances}},
  \href{https://doi.org/10.1051/0004-6361/201423496}{\emph{\aap} {\bfseries
  570} (2014) A13}.

\bibitem{Cosmicflows3}
R.~B. {Tully}, H.~M. {Courtois} and J.~G. {Sorce}, \emph{{Cosmicflows-3}},
  \href{https://doi.org/10.3847/0004-6256/152/2/50}{\emph{\aj} {\bfseries 152}
  (2016) 50}.

\bibitem{EDD09}
R.~B. {Tully}, L.~{Rizzi}, E.~J. {Shaya}, H.~M. {Courtois}, D.~I. {Makarov} and
  B.~A. {Jacobs}, \emph{{The Extragalactic Distance Database}},
  \href{https://doi.org/10.1088/0004-6256/138/2/323}{\emph{\aj} {\bfseries 138}
  (2009) 323}.

\bibitem{070809GCN}
D.~A. {Perley}, J.~S. {Bloom}, M.~{Modjaz}, A.~A. {Miller}, J.~{Shiode},
  J.~{Brewer} et~al., \emph{{GRB 070809: putative host galaxy and redshift.}},
  {\emph{GRB Coordinates Network} {\bfseries 7889} (2008) }.

\bibitem{helou1991nasa}
G.~Helou, B.~Madore, M.~Schmitz, M.~Bicay, X.~Wu and J.~Bennett, \emph{The
  nasa/ipac extragalactic database},  in \emph{Databases \& On-Line Data in
  Astronomy}, pp.~89--106, Springer, (1991).

\bibitem{090417GCN}
P.~T. {O'Brien} and N.~R. {Tanvir}, \emph{{GRB 090417A: nearby galaxy
  redshift.}}, {\emph{GRB Coordinates Network} {\bfseries 9136} (2009) }.

\bibitem{tunnicliffe2013nature}
R.~Tunnicliffe, A.~J. Levan, N.~R. Tanvir, A.~Rowlinson, D.~Perley, J.~Bloom
  et~al., \emph{On the nature of the ‘hostless’ short grbs}, {\emph{Monthly
  Notices of the Royal Astronomical Society} {\bfseries 437} (2013) 1495}.

\bibitem{tanvir2005origin}
N.~R. Tanvir, R.~Chapman, A.~J. Levan and R.~Priddey, \emph{An origin in the
  local universe for some short $\gamma$-ray bursts}, {\emph{\nat} {\bfseries
  438} (2005) 991}.

\bibitem{Fletcher2019gcn}
C.~{Fletcher}, \emph{{LIGO/Virgo S190425z: Fermi GBM Observations}}, {\emph{GRB
  Coordinates Network} {\bfseries 24185} (2019) }.

\bibitem{Carillo2019gcn}
A.~{Martin-Carillo}, V.~{Savchenko}, C.~{Ferrigno}, J.~{Rodi }, A.~{Coleiro}
  and S.~{Mereghetti}, \emph{{LIGO/Virgo S190425z: INTEGRAL prompt
  observation}}, {\emph{GRB Coordinates Network} {\bfseries 24169} (2019) }.

\end{thebibliography}\endgroup

\end{document}